\documentclass[12pt,a4paper]{article}
\pdfoutput=1
\usepackage{amsmath}
\usepackage{epsfig}
\usepackage{array}
\usepackage{float}
\usepackage{lscape,graphicx} 
\usepackage{graphics}
\usepackage{amssymb}
\usepackage{color}
\usepackage{multirow}
\usepackage{advdate}
\usepackage{datenumber}
\usepackage{ mathrsfs }
\usepackage{slashed}
\usepackage[utf8]{inputenc}

\newcommand{\gev}{\mbox{${\rm GeV}$}}
\newcommand{\bea}{\begin{equation}\begin{array}{c}}
\newcommand{\eea}{\end{array}\end{equation}}
\newcommand{\ea}{\end{array}}

\newcommand{\beq}{\begin{equation}}
\newcommand{\eeq}{\end{equation}}
\newcommand{\bad}{\begin{array}{ccc}}

\newcommand{\ba}{\begin{array}{c}}

\newcommand{\mdm}{M_{\chi_1}}
\newcommand{\mtwo}{M_{\chi_2}}
\newcommand{\tev}{\mathrm{TeV}}

\newcommand{\md}{M_D}
\newcommand{\ms}{M_S}
\newcommand{\tl}{\theta_L}
\newcommand{\tr}{\theta_R}
\newcommand{\sdm}{\mathrm{SD^3M}}

\begin{document}

\title{
{\bf Singlet-Doublet Dirac Dark Matter}}
\author{Carlos E. Yaguna$\footnote{carlos.yaguna@mpi-hd.mpg.de}$\\[7mm]
\it   Max-Planck-Institute f\"ur Kernphysik,\\
\it \small Saupfercheckweg 1, 69117 Heidelberg, Germany
}
\date{}
\maketitle
\thispagestyle{empty}
\begin{abstract}
We analyze a simple extension of the Standard Model where the dark matter particle is a Dirac fermion that is mixture of a singlet and  an $SU(2)$ doublet. The model contains only four free parameters: the singlet and the doublet masses and two new Yukawa couplings. Direct detection bounds in this model are very strong and require  the dark matter particle to be singlet-like. As a result, its relic density has to be obtained via coannihilations with the doublet. We find that the dark matter mass should be below $750$ GeV, that the singlet-doublet mass difference cannot exceed $9\%$, and that direct detection experiments offer the best chance to probe this scenario. Finally, we also show that this model can effectively arise in well-motivated  extensions of the Standard Model. 
\end{abstract}
\newpage
\section{Introduction}
The existence of dark matter provides clear evidence for physics beyond the Standard Model (SM) but does not tell us what this new physics should be. A direct approach to this problem is to construct minimal or simplified dark matter models. That is, to modify or extend the SM in simple ways that allow to explain the dark matter. Such models typically contain a small number of new fields and a reduced parameter space,  making them predictive and amenable to detailed analyses. Additionally, such minimal models can often be seen as limited cases of more complicated and better motivated theories. 

One of these simplified scenarios is the Singlet-Doublet fermion model \cite{ArkaniHamed:2005yv,Mahbubani:2005pt,D'Eramo:2007ga,Enberg:2007rp}, in which the dark matter candidate is a mixture of a  singlet and an $SU(2)$  doublet.  In this model, the dark matter particle is  a Majorana fermion. Consequently,  its spin-independent scattering  through the Z boson vanishes, and only the Higgs-mediated diagram contributes. To achieve the observed dark matter density via thermal freeze-out, an appropriate degree of mixing between the singlet and the doublet is required. In certain regions of the parameter space, this model describes bino-Higgsino dark matter in the MSSM and singlino-Higgsino dark matter in the NMSSM. The phenomenology of the Singlet-Doublet fermion model has been extensively studied in the recent literature \cite{Cohen:2011ec,Cheung:2013dua,Restrepo:2015ura,Calibbi:2015nha}.   

We want to analyze instead a related model where the dark matter particle is also a mixture of a singlet and an $SU(2)$ doublet but it is a Dirac fermion. We have dubbed this model the Singlet-Doublet Dirac Dark Matter model, or $\sdm$ model  for short. In this model, the particle content of the SM is extended with two Dirac fields, a singlet and a doublet, and the resulting Lagrangian contains only 4 new parameters. To our knowledge, this simple scenario has not been investigated before. Its phenomenology is quite interesting and very different from that of the  Singlet-Doublet fermion model. Direct detection bounds, for instance, play a more prominent role because the dark matter, being  a Dirac particle, may now elastically scatter on nuclei via Z mediated processes. As we will show, direct detection constraints \cite{Aprile:2012nq,Akerib:2013tjd} force the dark matter particle to be essentially a singlet, with a very small doublet component. In consequence, the relic density constraint can only  be satisfied thanks to  coannihilations \cite{Griest:1990kh} with the doublet, leading to a degenerate spectrum. We will determine the viable parameter space  of the $\sdm$ model and analyze its detection prospects. In addition, we will show that this model can easily be embedded into well-motivated extensions of the Standard Model. 

The rest of the paper is organized as follows. In the next section we describe the model and introduce our notation. The dark matter phenomenology is qualitatively discussed in section \ref{sec:dm}. Section \ref{sec:num} presents our main results. We use a scan over the parameter of this model to determine the viable regions and to analyze the detection prospects. In section \ref{sec:ext} we show that the $\sdm$ model can effectively arise in well-motivated extensions of the Standard Model. Finally, we summarize our results and draw our conclusions in section \ref{sec:con}.

\section{The model}
\label{sec:mod}
The model we consider extends the SM particle content with two Dirac fermions: an $SU(2)$ doublet with $Y=-1/2$, $\psi_{L,R}$, and  a  singlet, $S_{L,R}$, that interact among themselves and with the SM fields via the following Lagrangian: 
\begin{align}
\mathscr{L}_{\sdm}=&i\bar\psi\slashed D\psi+ i\bar S\slashed \partial S- M_D\bar\psi_L\psi_R-M_S \bar S_LS_R-y_1\bar \psi_L\tilde H S_R-y_2\bar \psi_R\tilde H S_L+h.c.
\label{eq:lag}
\end{align}
where $M_{D,S}$  are mass parameters, $y_{1,2}$ are new Yukawa couplings, $H$ is the SM Higgs doublet, and $\tilde H=i\sigma_2H$. The vector-like character of the new fermion doublet ensures that this model is free of gauge anomalies.

The above Lagrangian possesses a $Z_2$ symmetry under which the SM particles are even while the new fermions are odd, guaranteeing the stability of the lightest new fermion --the would-be dark matter particle. Such a symmetry, however, also allows  Majorana masses for $S_{L,R}$,  which would  give rise to a dark matter particle of Majorana type and, therefore, to the previously studied Singlet-Doublet fermion model \cite{Cohen:2011ec,Cheung:2013dua,Restrepo:2015ura,Calibbi:2015nha}. In this paper we want to focus instead on Dirac Dark Matter,  so we need to find a way of preventing such Majorana mass terms in the Lagrangian. One possibility to do so is by postulating a global $U(1)$ symmetry under which the new fields have all the same charge while the SM fields are neutral. A potential drawback of this approach is that such global symmetry is expected to be broken by gravitational effects at the Planck scale, inducing dark matter decay and threatening the viability of the model. According to \cite{Mambrini:2015sia}, current bounds require the effective coupling of the Planck-suppressed non-renormalizable operator that induces the decay of the dark matter particle to be smaller than about  $10^{-8}$.  A more interesting alternative is to promote that $U(1)$ to a gauge symmetry under which the singlets are charged. Such theory will not be exactly described by our Lagrangian but there will be a region of the parameter space where both will effectively coincide. A well-motivated example of this type will be discussed in section \ref{sec:ext}. For the following two sections, we will simply take equation (\ref{eq:lag}) as the Lagrangian of our model.

Thus, even before imposing any constraint, the model contains only $4$ free parameters. This small and manageable  parameter space is one of the main advantages of this scenario. Our aim is to determine the viable regions within that parameter space and to analyze the detection prospects of this model.

After electroweak symmetry breaking, the Yukawa interactions induce mixing between the singlet and the neutral component of the doublet.  In the basis ($S$, $\psi$) the resulting neutral fermion mass matrix is given by 
\beq
M=\begin{pmatrix}
M_S & \frac{y_2 v}{\sqrt 2}\\
\frac{y_1 v}{\sqrt 2} & M_D\end{pmatrix}, 
\eeq
where $v=246~\gev$. This matrix can be diagonalized by a bi-unitary transformation such that
\beq
M^d=\begin{pmatrix}\mdm& 0\\0& \mtwo\end{pmatrix}=U_L^\dagger MU_R, 
\eeq
with
\beq
U_{L,R}=\begin{pmatrix}\cos\theta_{L,R} & \sin\theta_{L,R}\\-\sin\theta_{L,R} & \cos\theta_{L,R}\end{pmatrix}.
\eeq
The lightest of the two neutral fermions ($\chi_1$, $\chi_2$) is the dark matter candidate. As we will see in the next section, the dark matter particle should be mostly singlet and therefore corresponds to $\chi_1$.

The mixing angles, $\theta_{L,R}$, can be written in terms of the original parameters of the model as
\begin{align}
\tan 2\tl&=\frac{\sqrt2 v\,(M_S\, y_1+M_D\, y_2)}{M_D^2-M_S^2+\frac{v^2}{2}(y_1^2-y_2^2)}\label{eq:mix1}\\
\tan 2\tr&=\frac{\sqrt2 v\,(M_S\, y_2+M_D\, y_1)}{M_D^2-M_S^2+\frac{v^2}{2}(y_2^2-y_1^2)},\label{eq:mix2}
\end{align}
while the masses of the neutral fermions are given by
\begin{align}
M_{\chi_1}^2&=\frac{\cos^2\tl(M_S^2+y_2v^2/2)-\sin^2\tl(M_D^2+y_1^2v^2/2)}{\cos^2\tl-\sin^2\tl}\\
M_{\chi_2}^2&=\frac{\sin^2\tl(M_S^2+y_2v^2/2)-\cos^2\tl(M_D^2+y_1^2v^2/2)}{\sin^2\tl-\cos^2\tl}.
\end{align}
The physical free parameters of the $\sdm$ model are then $\mdm$, $\mtwo$, $\tl$, and $\tr$.

In addition to $\chi_1$ and $\chi_2$, the spectrum also contains a new charged fermion, $\chi^+$, with mass $M_{\chi^+}=M_D+341$ MeV. LEPII bounds constrain this mass to be larger than about $100~\gev$ \cite{Agashe:2014kda}. 

\section{Dark Matter}
\label{sec:dm}
\begin{figure}[t!]
\begin{center}
\includegraphics[width=0.6\textwidth]{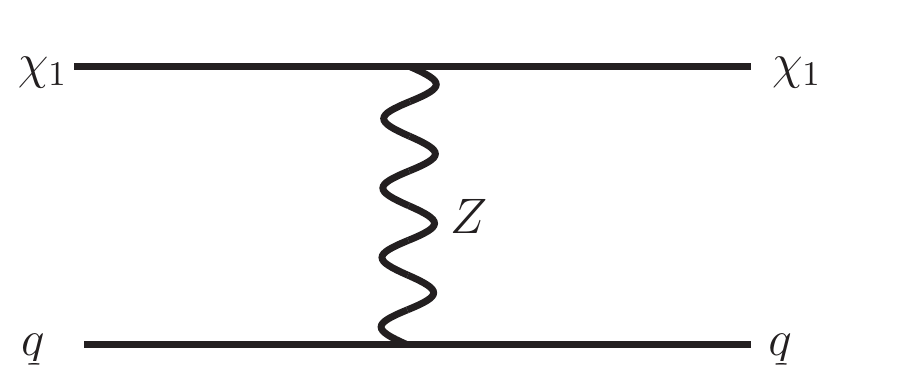} 
\caption{\small\it The Feynman diagram contributing to quark-dark matter scattering in this model. To be consistent with current direct detection bounds, the dark matter particle must be singlet-like so that its coupling to the $Z$ is mixing-suppressed. }
\label{fig:feyn}
\end{center}
\end{figure}
The novel feature of the $\sdm$ model is that the dark matter particle,  a mixture of the singlet and the doublet states,  is a Dirac fermion. In consequence, it has a non-zero vector coupling with the $Z$ that induces a spin-independent scattering with nuclei --see figure \ref{fig:feyn}. If the dark matter particle had a sizable doublet component, the resulting cross section would be orders of magnitude above present bounds. Thus, direct detection experiments require the dark matter particle to be mostly singlet: $\tl,\tr\ll1$ and $\mdm<\mtwo$. In that case, the spin-independent direct detection cross section with a nucleus can be written as 
\begin{equation}
\sigma_{SI}^{A,Z}=\frac{G_F^2 \mu^2}{8\pi}(\sin^2\tl+\sin^2\tr)^2 \left[(1-4\sin^2\theta_W)Z-(A-Z)\right]^2,
\end{equation}
where $\theta_W$ is the weak-mixing angle. Notice that $\sigma_{SI}^{A,Z}$ does not depend on the dark matter mass and is free of the uncertainties associated with the scalar matrix elements. Since $1-4\sin^2\theta_W$ is small, the dark matter particle hardly interacts with protons and $\sigma_{SI}^{A,Z}$ is essentially proportional to the dark matter-neutron cross section.  The relevant quantity to compare with the experimental bounds, however, is not $\sigma_{SI}^{A,Z}$ but the scattering cross section per nucleon \cite{Lewin:1995rx}, $\sigma_{SI}^N$, which is given by 
\beq
\label{eq:sip}
\sigma_{SI}^N=\frac{m_N^2}{\mu^2 A^2}\sigma_{SI}^{A,Z},
\eeq
where $m_N$ is the mass of the nucleon.  Compatibility with current direct detection limits requires $\sin\theta_{L,R}\lesssim 0.1-0.01$. Consequently, $\mdm=\ms$ and $\mtwo=\md$ to a very good approximation. 

Since the dark matter particle, $\chi_1$, is mostly singlet, it does not annihilate efficiently in the early Universe, with the result that its present abundance would normally exceed the observed value. Within the standard cosmological model, the only  way of avoiding this outcome and obtaining a relic density compatible with current data is via coannihilations with the doublet. Indeed, the doublet is known to have a large annihilation rate, yielding a thermal relic abundance for masses of order $1.1$ TeV. For $\mdm$  below that value, coannihilations between the singlet and the doublet may bring the dark matter density within the observed range. In fact, the largest possible value of  $\mdm$ consistent with this description can be easily estimated analytically.  Since the doublet annihilation cross section is much larger than the singlet one, when $\chi_1$ and $\chi_2$ are quasi-degenerate we get that (see e.g. \cite{Profumo:2004wk})
\begin{equation}
\Omega_{\chi_1}h^2=\Omega_{\chi_2}h^2\left(\frac{g_\psi+g_S}{g_\psi}\right)^2,
\end{equation}  
being $g_S=4$ and $g_\psi=8$ respectively  the number of degrees of freedom  for the singlet and the doublet (including its charged component). Given that $\Omega_{\chi_2}h^2\approx0.11\left(\frac{\mtwo}{1.1~\mathrm{TeV}}\right)^2$, and $\mdm\sim \mtwo$, we find that $\mdm\lesssim 733~\gev$.

To summarize, in this model the dark matter particle is a Dirac fermion that  is essentially singlet under the SM gauge group, obtains its relic density thanks to  coannihilations with the doublet, and interacts with nuclei via mixing angle-suppressed weak interactions.  In the next section, we will numerically study the parameter space of this model, analyze the resulting phenomenology and determine the dark matter detection prospects.   

\begin{figure}[t!]
\begin{center}
\begin{tabular}{cc}
\includegraphics[width=0.5\textwidth]{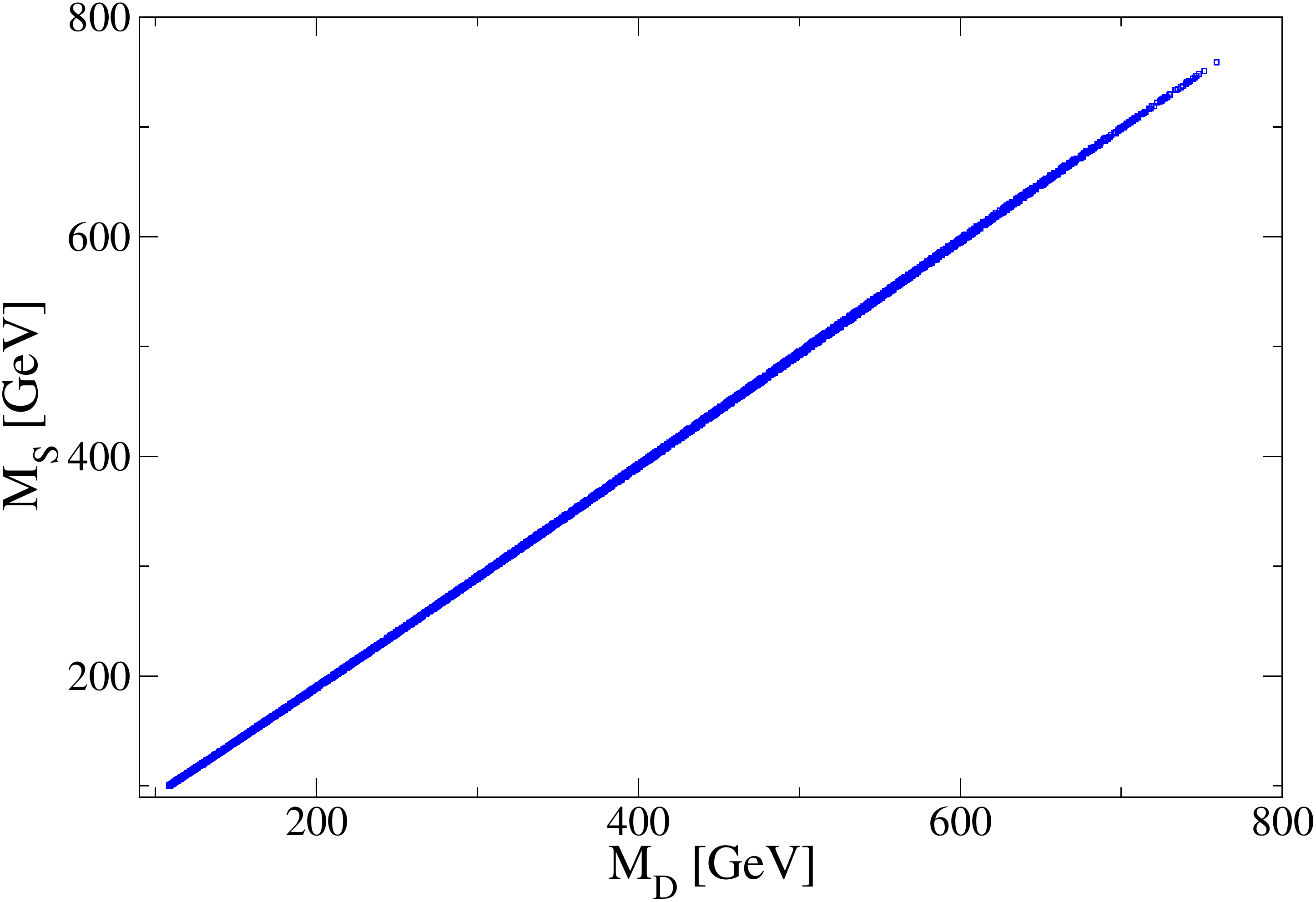}
& \includegraphics[width=0.5\textwidth]{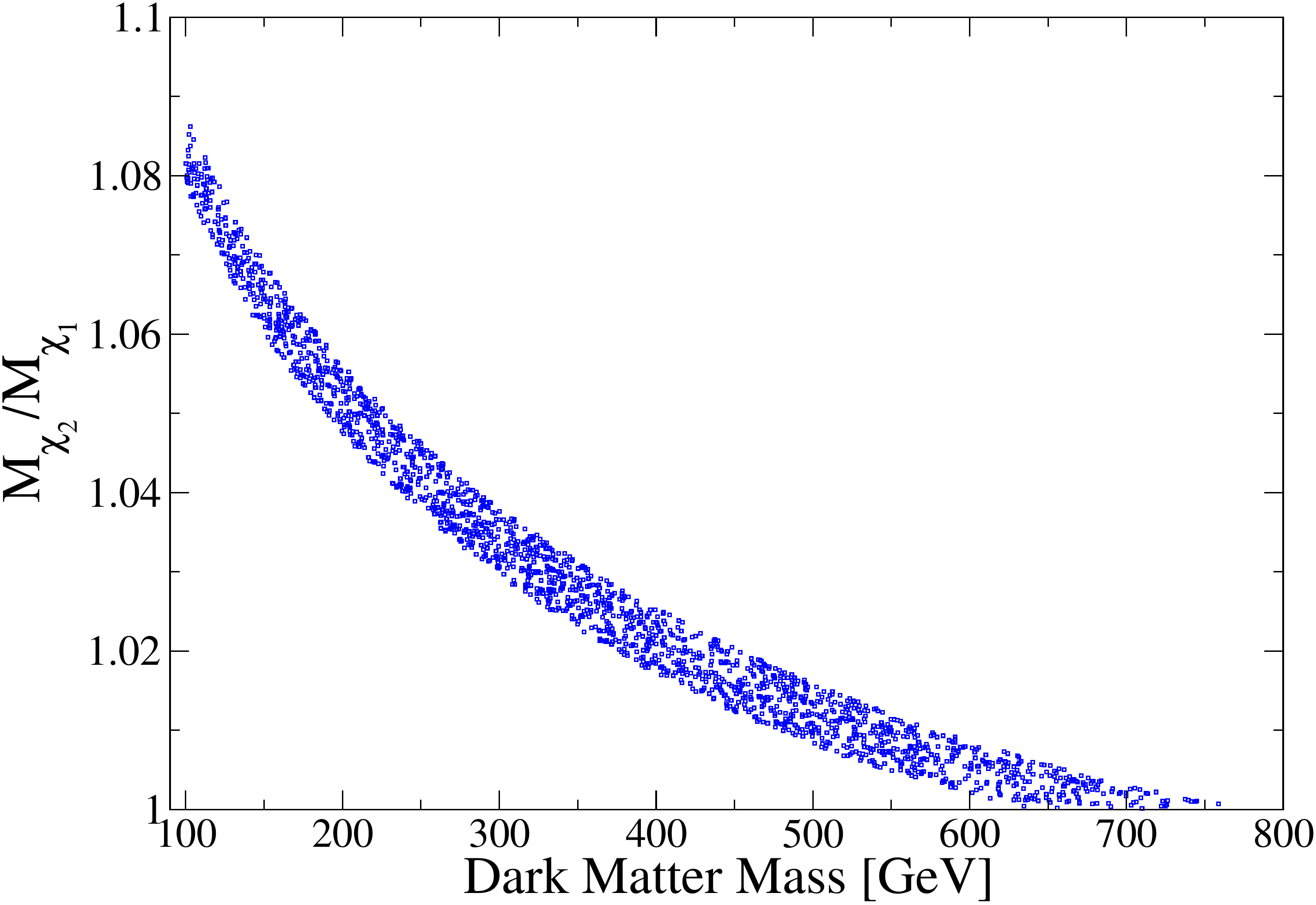}
\end{tabular} 
\caption{\small\it The viable models projected onto  the planes ($\md$, $\ms$) and ($\mdm$, $M_{\chi_2}/\mdm$). Since the relic density is determined by coannihilations, the mass splitting is always quite small: $\md\sim \ms$ and $\mtwo/\mdm\lesssim 1.1$.}
\label{fig:mass}
\end{center}
\end{figure}

\section{Numerical Results}
\label{sec:num}

We have scanned the parameter space of this model ($\md,\ms< 2~\tev$, $y_1,y_2>10^{-6}$) and obtained a large sample of viable models. All these viable models give a relic density in agreement with current observations \cite{Ade:2013zuv} and a spin-independent direct detection cross section consistent with the LUX bounds \cite{Akerib:2013tjd}. For the calculation of the relic density we have used micrOMEGAs \cite{Belanger:2013oya}(and LanHEP \cite{Semenov:2010qt}), which automatically takes into account all coannihilation processes. In this section we analyze that sample of viable models in some detail.

First of all, we found that the viable models satisfy $\mdm/\ms\approx 1$ and $\mtwo/\md\approx 1$ to a precision better than $10^{-4}$. Thus, in the following figures one can always replace  $\mdm$ ($\mtwo$) for $\ms$ ($\md$). We show the viable models projected onto the plane ($\md$, $\ms$) in the left panel of figure \ref{fig:mass}.  All viable points turn out to lie very close to  the line $\md=\ms$. Far from that region the points would have either a relic density not compatible with the observed dark matter density or a direct detection cross section much larger than allowed by current data.  From the figure we see that the upper bound on $\md,\ms$ is of order $750~\gev$, consistent with our estimate in the previous section. This upper bound is smaller than that  found in the Singlet-Doublet fermion model, where it reaches approximately $1.1$ TeV.  The right panel shows instead the ratio $\mtwo/\mdm$ --relevant for coannihilations-- versus the dark matter mass. This ratio always lies below $9\%$ and, as expected,  it decreases with the dark matter mass. For $\mdm=200, 500, 600 ~\gev$ the ratio $\mtwo/\mdm$ is respectively of order $6\%,2\%,1\%$. Close to $\mdm\sim 750~\gev$, the mass difference between $\chi_1$ and $\chi_2$ becomes negligible.

\begin{figure}[t!]
\begin{center}
\begin{tabular}{cc}
\includegraphics[width=0.5\textwidth]{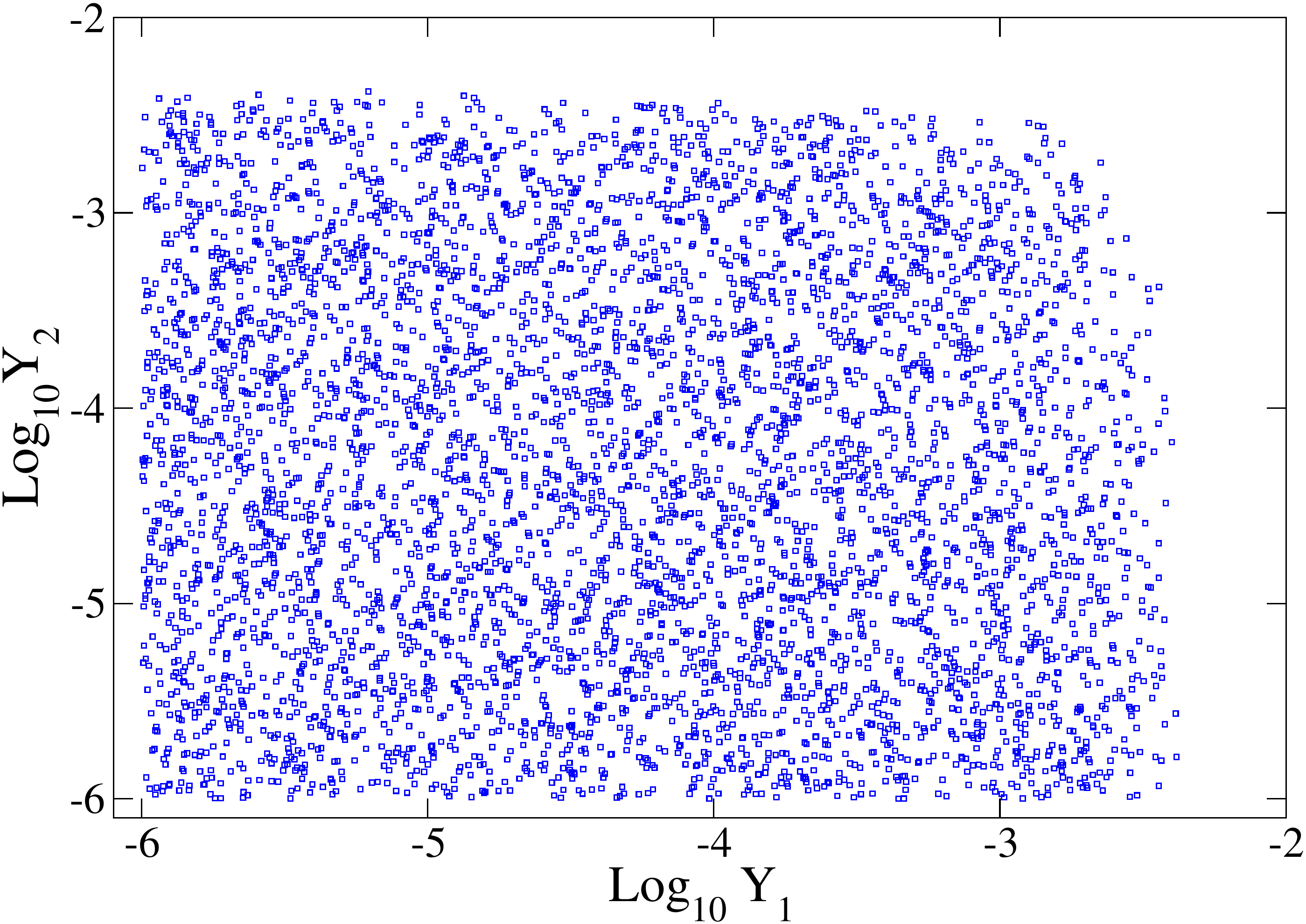}
& \includegraphics[width=0.5\textwidth]{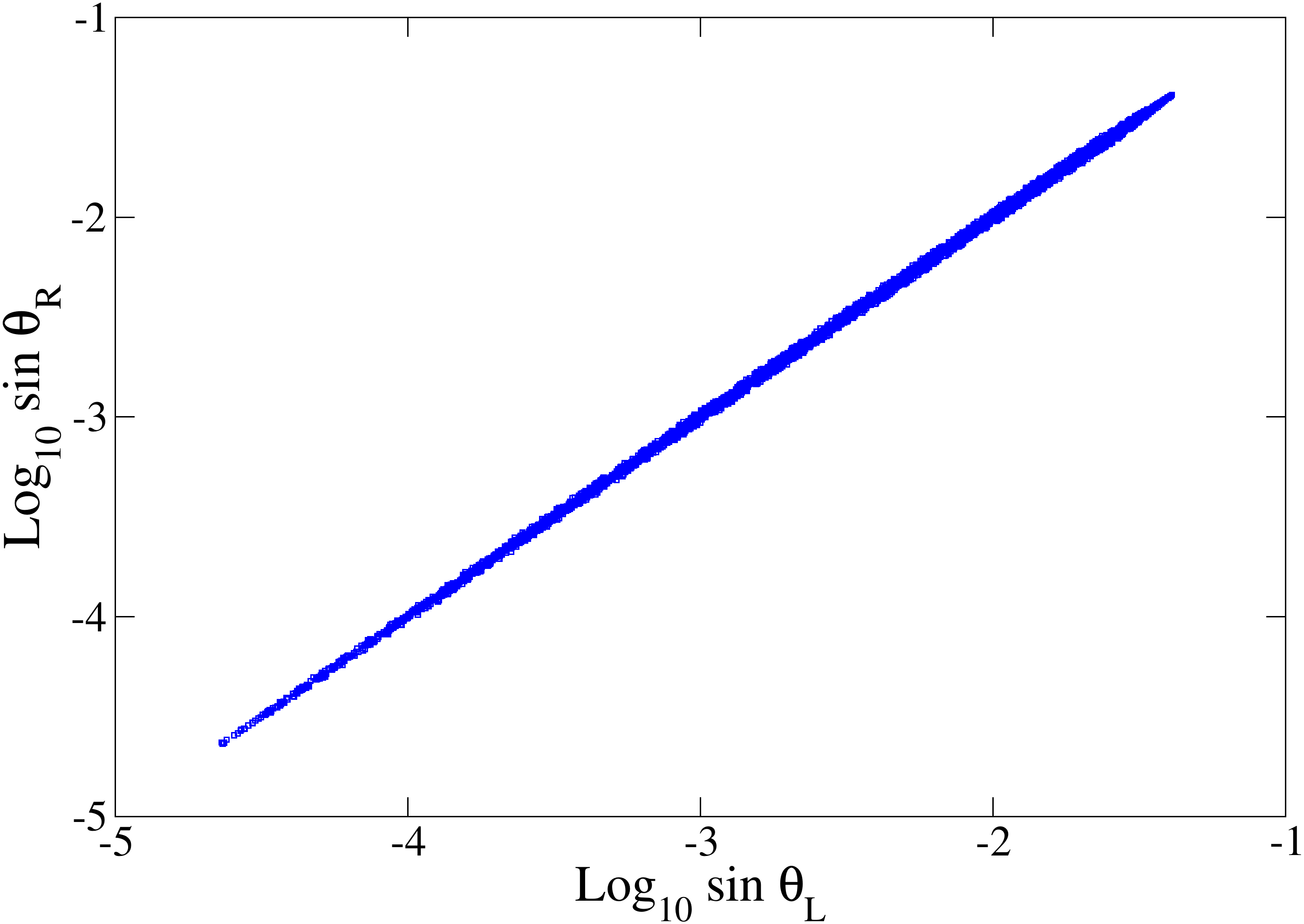}
\end{tabular} 
\caption{\small\it The viable models projected onto the planes ($y_1$, $y_2$) and ($\sin\theta_L$, $\sin\theta_R$). The upper bound on the Yukawas and on the mixing angles are determined by the direct detection bound from LUX.}
\label{fig:yuk}
\end{center}
\end{figure}

The mass spectrum consistent with the observed relic density and direct detection limits is, therefore, essentially degenerate, and consists of a dark matter particle with mass $\mdm=\ms$ and two slightly heavier fermions, $\chi^+,\chi_2$, with almost the same mass $\md$ ($\md=\ms+\Delta m$).  Such degenerate spectrum is very challenging for collider searches at the LHC.  The most sensitive searches are monojet signatures ($pp\to \chi_a\chi_bj$), which have a large background from $Z+jets$ and $W+jets$.  In \cite{Barducci:2015ffa}, it was found that a similar scenario (for collider purposes) --Natural Supersymmetry with low values of the $\mu$ parameter-- cannot be  constrained by current LHC data, and that the 13 TeV LHC could probe dark matter masses up to $250$ GeV. Based on those results, we can claim that only a small fraction of the viable parameter space of the $\sdm$ model will be probed by searches at the 13 TeV LHC. 

\begin{figure}[t!]
\begin{center}
\includegraphics[width=0.8\textwidth]{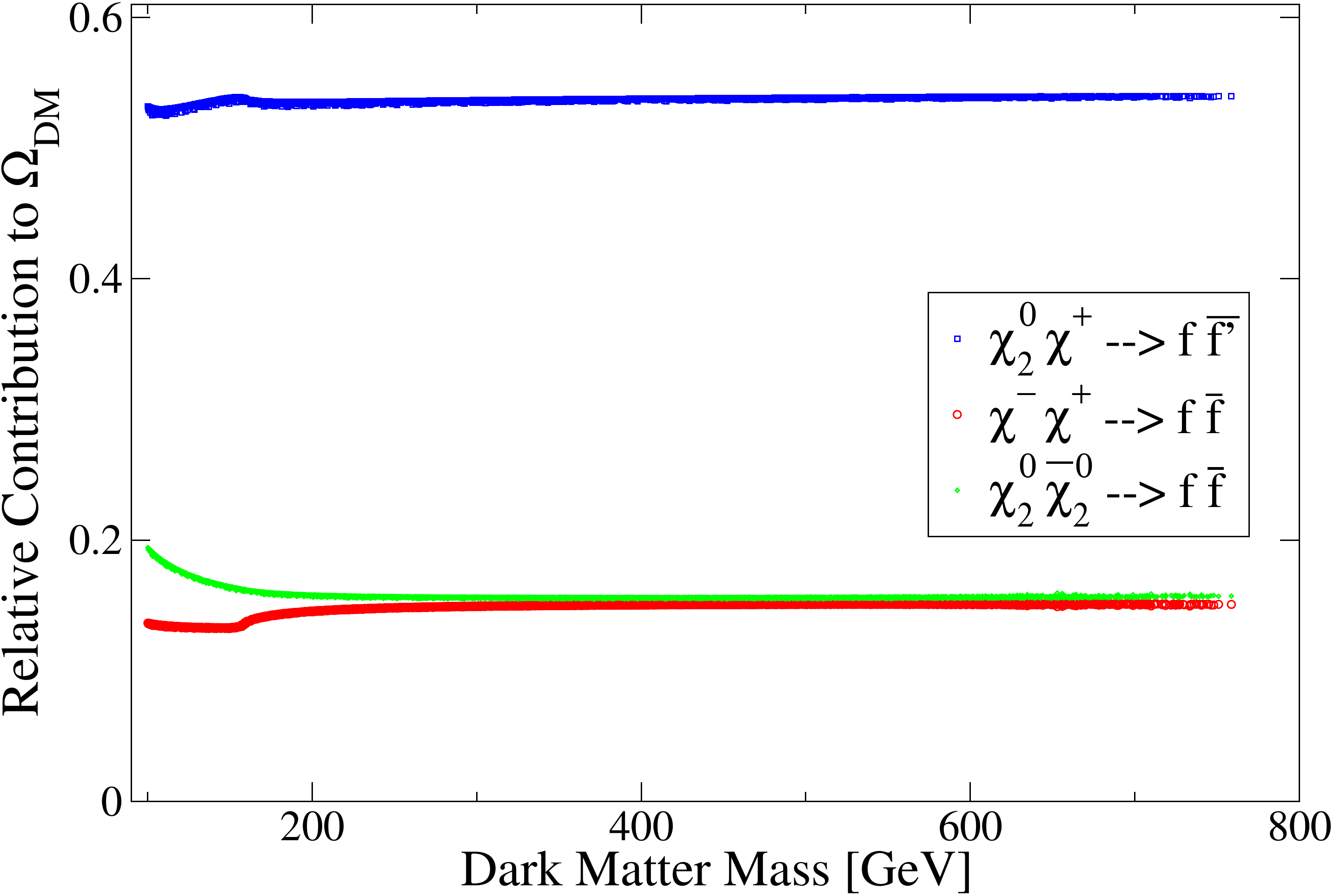} 
\caption{\small\it The dominant processes contributing to the dark matter annihilation rate in the early Universe. Notice that the dark matter particle, $\chi_1$, does not participate in these processes.}
\label{fig:coannch}
\end{center}
\end{figure}

The allowed values for the Yukawa couplings are illustrated in the left panel of figure \ref{fig:yuk}. As a result of the direct detection constraint, they fulfill  $y_1,y_2\lesssim 4\times10^{-3}$. Thus, the dark matter Yukawa couplings must necessarily be small in this scenario.  Such small Yukawa couplings guarantee that the  contribution from the new fermions to the electroweak precision parameters, in particular to $T$, remain tiny --$\Delta T\propto (y_1^2-y_2^2)^2$ \cite{Enberg:2007rp}.  Notice that below their upper bound, the Yukawa couplings can take essentially any value. That is, the dark matter constraint does not restrict the possible values of $y_1$ and $y_2$.  The right panel of figure \ref{fig:yuk} shows the viable points projected onto the plane ($\sin\theta_L$,  $\sin\theta_R$). Due to  $y_1,y_2\ll1$ and $\md\sim \ms$, both mixing angles must be very similar, in agreement with equations (\ref{eq:mix1}) and (\ref{eq:mix2}). In fact, the ratio $\sin\theta_L/\sin\theta_R$ is very close to $1$ at large dark matter masses and it varies between $0.92$ and $1.08$ for  $\mdm\sim 100~\gev$.  The upper bound on $\sin\theta_{L,R}$ observed in the figure ($\sim 4\times 10^{-2}$) is determined by the direct detection limit from LUX, according to equation (\ref{eq:sip}).

In this setup, the dark matter relic density is entirely determined by coannihilation processes, as illustrated in figure \ref{fig:coannch}. It displays the contribution of the different annihilation processes to the total dark matter annihilation rate in the early Universe. The dominant process, accounting for about $55\%$ of the total rate,  is the annihilation of the charged fermion and the heavier neutral fermion into SM fermions mediated by a $W^\pm$,  $\chi_2^0\chi^+\to f\bar f'$ (blue points). The processes $\chi^+\chi^-\to f\bar f$ (green points) and $\chi_2\bar\chi_2\to f\bar f$ (red points) together contribute another $30\%$ or so. The remaining $15\%$ (not shown in the figure) is accounted for by annihilation into gauge bosons such as $\chi_2\chi^+\to A/Z\,W^+$,  $\chi^+\chi^-\to W^+W^-$ and $\chi_2\bar\chi_2\to ZZ$. Processes involving the annihilation of dark matter particles give a negligible contribution because they are suppressed by the Yukawa couplings or the mixing angles.

\begin{figure}[t!]
\begin{center}
\includegraphics[width=0.8\textwidth]{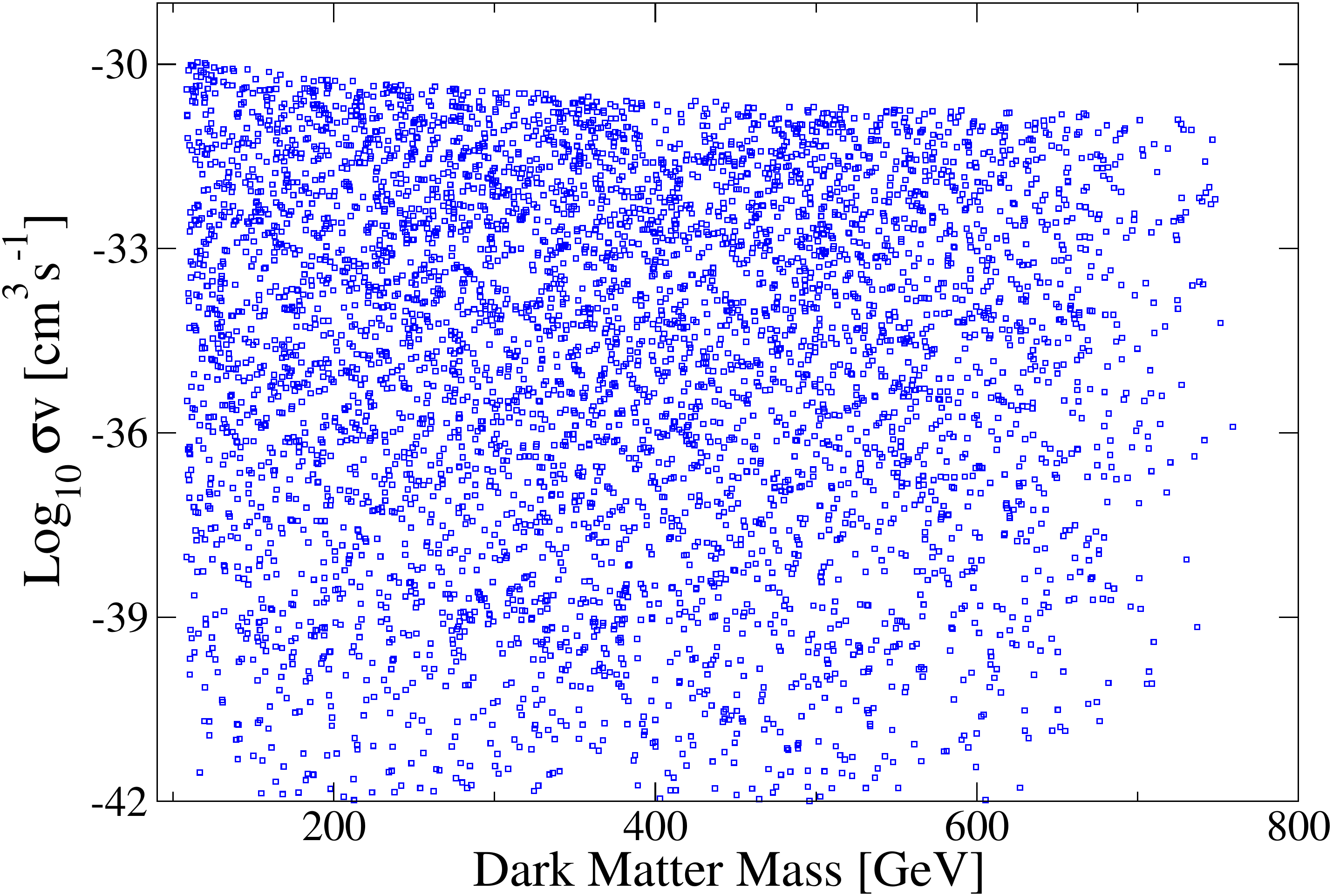} 
\caption{\small\it The present dark matter annihilation rate, $\sigma v$, versus the dark matter mass for the viable models. Since the relic density is obtained via coannihilatios, $\sigma v$ lies orders of magnitude below the so-called thermal value. No indirect detection signals are expected in this model.}
\label{fig:sigmav}
\end{center}
\end{figure}
Figure \ref{fig:sigmav} shows the annihilation rate today, $\sigma v$, versus the dark matter mass. Since the relic density is obtained via coannihilations, $\sigma v$ turns out to be much smaller than the so-called thermal value ($3\times 10^{-26}\mathrm{cm^3s^{-1}}$), lying instead below $10^{-30}\mathrm{cm^3s^{-1}}$. In fact, annihilation processes such as $\chi_1\bar \chi_1\to f\bar f, W^+W^-$ are strongly suppressed by Yukawa couplings and mixing angles. In consequence, no observable indirect detection signal is expected in this model. Conversely, if a signal were confirmed in any indirect detection experiment, we could immediately exclude the $\sdm$ model.

\begin{figure}[t!]
\begin{center}
\includegraphics[width=0.8\textwidth]{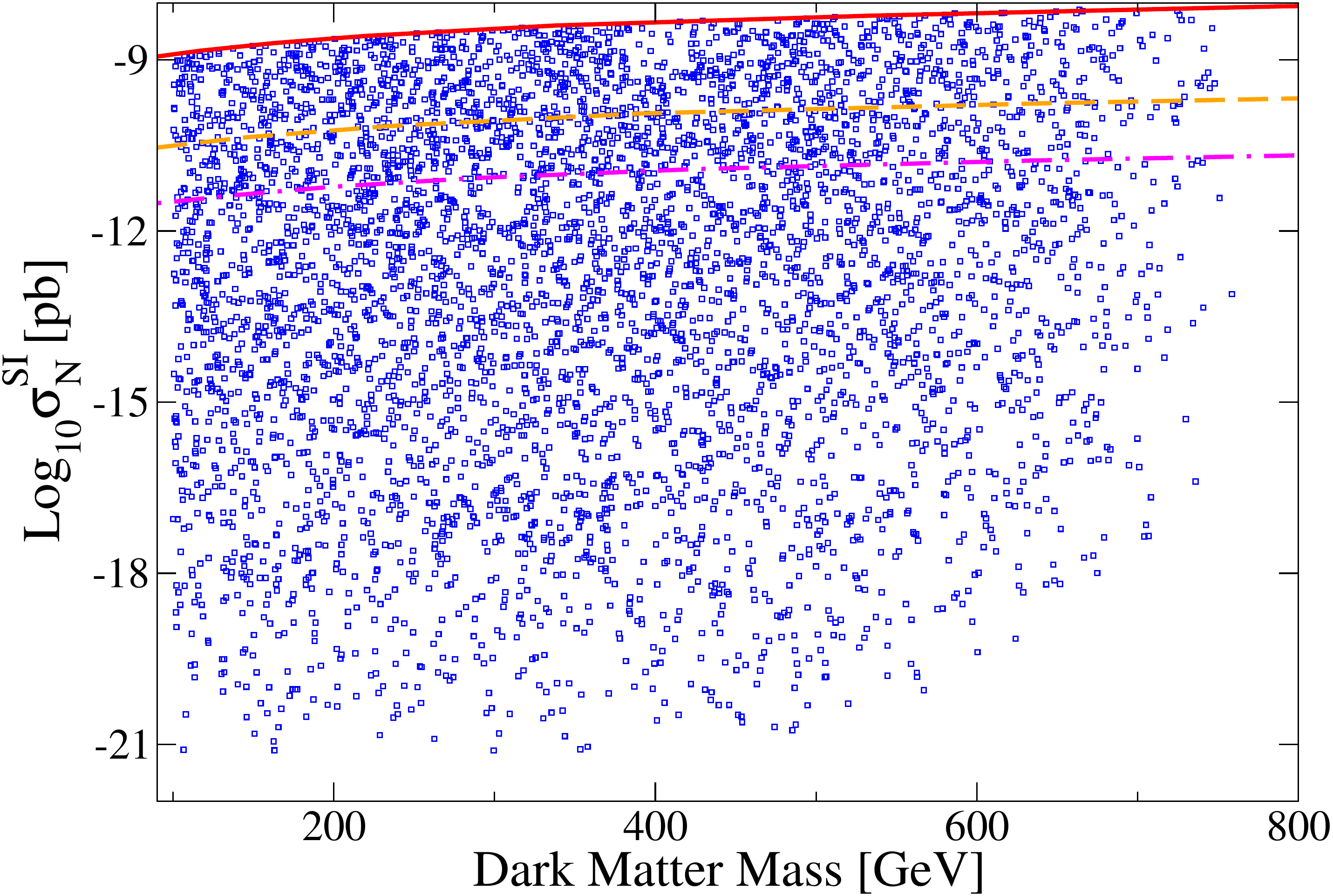} 
\caption{\small\it The spin-independent dark matter-nucleon cross section versus the dark matter mass for the viable models. The variation in $\sigma^{SI}_N$ is due to the mixing angles --see equation (\ref{eq:sip}). The red solid line shows the current experimental bound from LUX. The other two lines correspond to   the expected sensitivities of XENON1T (orange dashed line) and LZ (magenta dash-dotted line).}
\label{fig:ddsi}
\end{center}
\end{figure}

More promising is the possibility of testing this model via direct detection experiments. Figure \ref{fig:ddsi} displays the predicted spin-independent direct detection cross section as a function of the dark matter mass. By construction, all the points are consistent with the current bound from LUX \cite{Akerib:2013tjd} (solid red line). For comparison, we also show the expected sensitivities of XENON1T \cite{Aprile:2012zx} (dashed orange line) and LZ \cite{Malling:2011va} (dash-dotted magenta line).  Since the  spin-independent cross section strongly depends on the mixing angles, see equation (\ref{eq:sip}), and they  can in principle vary over a wide range, the allowed values of $\sigma^{SI}_N$ span more than $10$ orders of magnitude. As can be seen in the figure, many models feature cross sections within the expected sensitivity of current and planned experiments. If a direct detection signal were indeed observed in the near future, one could derive the dark matter mass and the mixing angles directly from the measurement, using equation (\ref{eq:sip}). Then,  the relic density constraint can be used, via figure \ref{fig:mass}, to estimate the mass difference. In this way, one could hope to eventually reconstruct the parameters of the model. 

The spin-dependent scattering of the dark matter particle with a nucleus also proceeds through a $Z$ mediated diagram, but it always has a small cross section. We found that the viable models feature $\sigma_{SD}^N<10^{-10}$ pb, well below the expected sensitivity of future experiments.

\section{$\boldsymbol{\sdm}$ in Gauge Extensions of the SM}
\label{sec:ext}

In the previous sections we took the Lagrangian in equation (\ref{eq:lag}) as our starting point, and studied in detail the resulting dark matter phenomenology. Now, we would like to demonstrate that such a Lagrangian can effectively arise in gauge extensions of the Standard Model.  

At first sight, the presence of the singlet fermions $S_{L,R}$ seems to be incompatible with a Dirac dark matter particle, because the Majorana mass terms $\overline{{S_{L,R}^c}}S_{L,R}$ are allowed,  giving rise to Majorana mass eigenstates and, in particular, to a Majorana dark matter particle. To obtain Dirac dark matter, we must find a way to forbid such Majorana mass terms. As already mentioned in section \ref{sec:mod}, one way to do so is to assume the existence of an additional $U(1)$ gauge symmetry under which $S_{L,R}$ are charged. The charges for the different fields can then be chosen to ensure that all the terms in (\ref{eq:lag}) be allowed and that the dark matter particle be automatically stable. If necessary, additional fermions would be introduced to cancel the anomalies. Such model would be effectively described by the Lagrangian in (\ref{eq:lag}) when the additional fermions are much heavier than the singlet and the doublet, and when the effect of the new gauge interaction is negligible (due to a small gauge coupling or a heavy gauge boson). An explicit and  well-motivated example of this framework is provided by the model with gauged baryon number \cite{Duerr:2013dza,Duerr:2013lka,Duerr:2014wra}.   

In this model, based on the gauge group $SU(3)\times SU(2)\times U(1)_Y\times U(1)_B$, baryon number ($B$) is promoted to a local symmetry that is spontaneously broken at a low scale. A realistic and anomaly-free realization of this theory  contains three vector-like fermions: an $SU(2)$ doublet with $Y=-1/2$, an $SU(2)$ singlet with $Y=-1$, and a SM singlet, all of them charged under $U(1)_B$. The particle content of this model thus includes the singlets and doublets of the $\sdm$ model. The model also contains an additional scalar that spontaneously breaks baryon number and generates vector-like masses for the new fermions. That is, in this model the parameters $M_D$ and $M_S$ in equation (\ref{eq:lag}) are actually proportional to the baryon number breaking scale. This model satisfies the properties mentioned above, so there is a region of the parameter space where its dark matter phenomenology is effectively described by the $\sdm$ model. This region of the parameter space was not studied in \cite{Duerr:2013lka,Duerr:2014wra}, where the effect of the Yukawa couplings --$y_{1,2}$ in (\ref{eq:lag})-- was neglected.

\section{Conclusions}
\label{sec:con}
We studied the Singlet-Doublet Dirac Dark Matter ($\sdm$) model, a minimal extension of the Standard Model featuring as a dark matter particle a Dirac fermion that is a mixture of a singlet and an $SU(2)$ doublet. The spectrum consists of three Dirac fermions, one charged and two neutrals, the lightest one being the dark matter particle. This scenario is very simple as it only contains four free parameters: the singlet and doublet masses ($\ms$, $\md$), and two Yukawa couplings ($y_1$, $y_2$). Alternative, one can take as free parameters the physical masses ($\mdm$, $\mtwo$) and the two mixing angles ($\tl$, $\tr$). Due to the strong bounds from direct detection experiments, the dark matter particle, $\chi_1$, has to be singlet-like, with a very small doublet component ($\tl,\tr\ll1$). The only way for such a particle to obtain a thermal relic density consistent with the observations is via coannihilations with the doublet. We found that, as a result, the dark matter mass should be below $750$ GeV and the singlet-doublet mass splitting cannot exceed $9\%$. The viable  spectrum is thus quite degenerate and very challenging for LHC searches. The dark matter density was  shown to be set  by the annihilation of the doublet components,  mostly $\chi_2\chi^+\to f\bar f'$, with a subdominant contribution from  $\chi^-\chi^+\to f\bar f$ and  $\chi_2\bar\chi_2\to f\bar f$. We also analyzed the dark matter detection prospects of this simplified scenario. Since  $\sigma v$ lies at least four orders of magnitude below the thermal value, the model predicts that no indirect detection signals will be found in the near future. Regarding direct detection, planned experiments will be able to probe new viable regions of the parameter space. Finally, we also demonstrated that the $\sdm$ model can effectively arise in well-motivated extensions of the Standard Model.    

\section*{Acknowledgments}
 I am supported by the Max Planck Society in the project MANITOP.

\bibliographystyle{hunsrt}
\bibliography{darkmatter}

\begin{thebibliography}{10}

\bibitem{ArkaniHamed:2005yv}
Nima Arkani-Hamed, Savas Dimopoulos, and Shamit Kachru.
\newblock {Predictive landscapes and new physics at a TeV}.
\newblock 2005, hep-th/0501082.

\bibitem{Mahbubani:2005pt}
Rakhi Mahbubani and Leonardo Senatore.
\newblock {The Minimal model for dark matter and unification}.
\newblock {\em Phys. Rev.}, D73:043510, 2006, hep-ph/0510064.

\bibitem{D'Eramo:2007ga}
Francesco D'Eramo.
\newblock {Dark matter and Higgs boson physics}.
\newblock {\em Phys. Rev.}, D76:083522, 2007, 0705.4493.

\bibitem{Enberg:2007rp}
R.~Enberg, P.~J. Fox, L.~J. Hall, A.~Y. Papaioannou, and M.~Papucci.
\newblock {LHC and dark matter signals of improved naturalness}.
\newblock {\em JHEP}, 11:014, 2007, 0706.0918.

\bibitem{Cohen:2011ec}
Timothy Cohen, John Kearney, Aaron Pierce, and David Tucker-Smith.
\newblock {Singlet-Doublet Dark Matter}.
\newblock {\em Phys. Rev.}, D85:075003, 2012, 1109.2604.

\bibitem{Cheung:2013dua}
Clifford Cheung and David Sanford.
\newblock {Simplified Models of Mixed Dark Matter}.
\newblock {\em JCAP}, 1402:011, 2014, 1311.5896.

\bibitem{Restrepo:2015ura}
Diego Restrepo, Andrés Rivera, Marta Sánchez-Peláez, Oscar Zapata, and
  Walter Tangarife.
\newblock {Radiative neutrino masses in the singlet-doublet fermion dark matter
  model with scalar singlets}.
\newblock {\em Phys. Rev.}, D92(1):013005, 2015, 1504.07892.

\bibitem{Calibbi:2015nha}
Lorenzo Calibbi, Alberto Mariotti, and Pantelis Tziveloglou.
\newblock {Singlet-Doublet Model: Dark matter searches and LHC constraints}.
\newblock 2015, 1505.03867.

\bibitem{Aprile:2012nq}
E.~Aprile et~al.
\newblock {Dark Matter Results from 225 Live Days of XENON100 Data}.
\newblock {\em Phys.Rev.Lett.}, 109:181301, 2012, 1207.5988.

\bibitem{Akerib:2013tjd}
D.S. Akerib et~al.
\newblock {First results from the LUX dark matter experiment at the Sanford
  Underground Research Facility}.
\newblock 2013, 1310.8214.

\bibitem{Griest:1990kh}
Kim Griest and David Seckel.
\newblock {Three exceptions in the calculation of relic abundances}.
\newblock {\em Phys.Rev.}, D43:3191--3203, 1991.

\bibitem{Mambrini:2015sia}
Yann Mambrini, Stefano Profumo, and Farinaldo~S. Queiroz.
\newblock {Dark Matter and Global Symmetries}.
\newblock 2015, 1508.06635.

\bibitem{Agashe:2014kda}
K.~A. Olive et~al.
\newblock {Review of Particle Physics}.
\newblock {\em Chin. Phys.}, C38:090001, 2014.

\bibitem{Lewin:1995rx}
J.~D. Lewin and P.~F. Smith.
\newblock {Review of mathematics, numerical factors, and corrections for dark
  matter experiments based on elastic nuclear recoil}.
\newblock {\em Astropart. Phys.}, 6:87--112, 1996.

\bibitem{Profumo:2004wk}
S.~Profumo and C.~E. Yaguna.
\newblock {Gluino coannihilations and heavy bino dark matter}.
\newblock {\em Phys. Rev.}, D69:115009, 2004, hep-ph/0402208.

\bibitem{Ade:2013zuv}
P.~A.~R. Ade et~al.
\newblock {Planck 2013 results. XVI. Cosmological parameters}.
\newblock {\em Astron. Astrophys.}, 571:A16, 2014, 1303.5076.

\bibitem{Belanger:2013oya}
G.~Belanger, F.~Boudjema, A.~Pukhov, and A.~Semenov.
\newblock {micrOMEGAs3.1: a program for calculating dark matter observables}.
\newblock 2013, 1305.0237.

\bibitem{Semenov:2010qt}
A.~Semenov.
\newblock {LanHEP - a package for automatic generation of Feynman rules from
  the Lagrangian. Updated version 3.1}.
\newblock 2010, 1005.1909.

\bibitem{Barducci:2015ffa}
Daniele Barducci, Alexander Belyaev, Aoife K.~M. Bharucha, Werner Porod, and
  Veronica Sanz.
\newblock {Uncovering Natural Supersymmetry via the interplay between the LHC
  and Direct Dark Matter Detection}.
\newblock {\em JHEP}, 07:066, 2015, 1504.02472.

\bibitem{Aprile:2012zx}
Elena Aprile.
\newblock {The XENON1T Dark Matter Search Experiment}.
\newblock 2012, 1206.6288.

\bibitem{Malling:2011va}
D.~C. Malling et~al.
\newblock {After LUX: The LZ Program}.
\newblock 2011, 1110.0103.

\bibitem{Duerr:2013dza}
Michael Duerr, Pavel Fileviez~Perez, and Mark~B. Wise.
\newblock {Gauge Theory for Baryon and Lepton Numbers with Leptoquarks}.
\newblock {\em Phys. Rev. Lett.}, 110:231801, 2013, 1304.0576.

\bibitem{Duerr:2013lka}
Michael Duerr and Pavel Fileviez~Perez.
\newblock {Baryonic Dark Matter}.
\newblock {\em Phys. Lett.}, B732:101--104, 2014, 1309.3970.

\bibitem{Duerr:2014wra}
Michael Duerr and Pavel Fileviez~Perez.
\newblock {Theory for Baryon Number and Dark Matter at the LHC}.
\newblock {\em Phys. Rev.}, D91(9):095001, 2015, 1409.8165.

\end{thebibliography}

\end{document}